\documentclass[summary]{URSIGASS2020}

\title{TCAD modeling for SiGe HBT THz detectors}

\author{Xueqing Liu\affref{ref1}, John Suarez\affref{ref2},
  and Michael Shur\affref{ref1}}

\affiliation{%
  \aff{ref1}{Department of Electrical, Computer, and Systems Engineering, Rensselaer Polytechnic Institute, Troy, NY 12180, USA}
  \aff{ref2}{Department of Electrical Engineering, Widener University, Chester, PA 19013, USA}
}

\begin{document}

\maketitle

\begin{abstract}
  Terahertz (THz) response of transistor and integrated circuit yields important information about device parameters and has been used for distinguishing between working and defective transistors and circuits. Using a TCAD model for SiGe HBTs we simulate their current-voltage characteristics and their response to sub-THz (300\,GHz) radiation. Applying different mixed mode schemes in TCAD, we simulated the dynamic range of the THz response for SiGe HBTs and showed that it is comparable with that of the TeraFET detectors. The HBT response to the variations of the detector design parameters are investigated at different frequencies with the harmonic balance simulation in TCAD. These results are useful for the physical design and optimization for the HBT THz detectors and for the identification of faulty SiGe HBT and Si BiCMOS circuits using sub-THz or THz scanning.
\end{abstract}

\section{Introduction}

Emerging beyond 5G communications and many other sub-THz and THz applications, communications, scanning, testing, biological, medical, and industrial control have stimulated the demand for efficient sub-THz and THz detectors~\cite{Akkas201901,Klatt200912,Federici200506,Cheon201611,Xie201503,Ahi201803}. Plasmonic TeraFETs implemented in InGaAs, GaAs, GaN, and Si have demonstrated efficient detection based on the excitation of decayed plasm waves~\cite{Dyakonov199605,Dyakonov199610,Popov201104,Gavrilenko200702,Knap200407}. InP HBTs have also demonstrated a reasonable performance in detecting THz radiation~\cite{Coquillat201612,Dyakonova2017}. SiGe HBTs and BiCMOS circuits, which are compatible with standard VLSI technology, have shown advantages for many applications ~\cite{Mai201706,Schroter201601,Hadi2012,Ghodsi201906}.

In this work, we report on the modeling of the SiGe HBTs for terahertz detection. The dynamic range of the SiGe HBT response has been simulated and compared with various TeraFET detectors. The SiGe HBT response to the variations of the device feature size ranging from 20\,nm to 130\,nm and other design parameters are explored for such detectors. These results could be important for testing Si BiCMOS VLSI using THz scanning.

\begin{figure}[t]
  \centering
  \includegraphics[width=0.85\columnwidth]{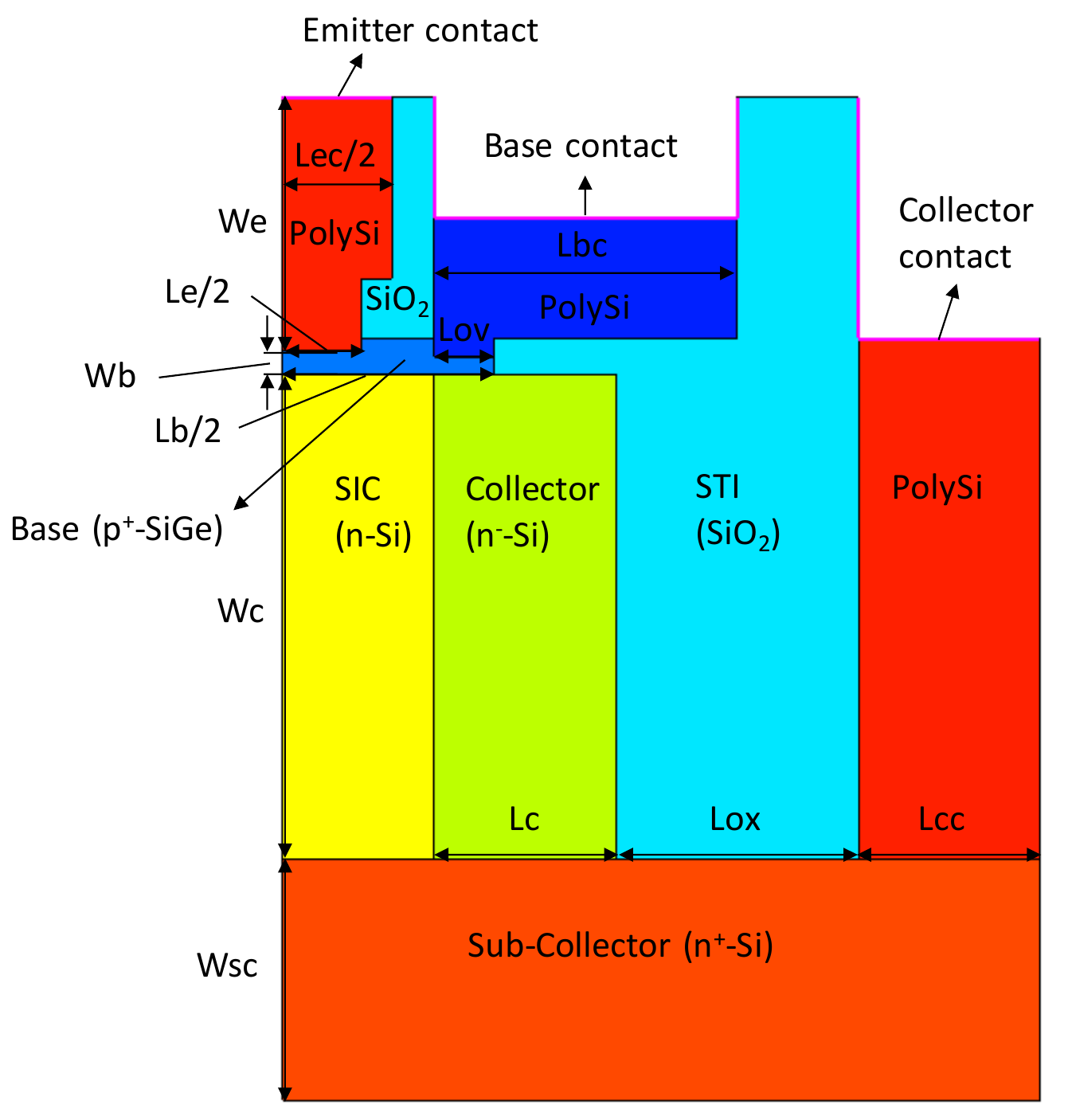}
  \caption{Schematic of the SiGe HBT structure in TCAD.}
  \label{fig1}
\end{figure}

\section{Model setup}

\begin{figure}[!ht]
  \centering
  \includegraphics[width=0.75\columnwidth]{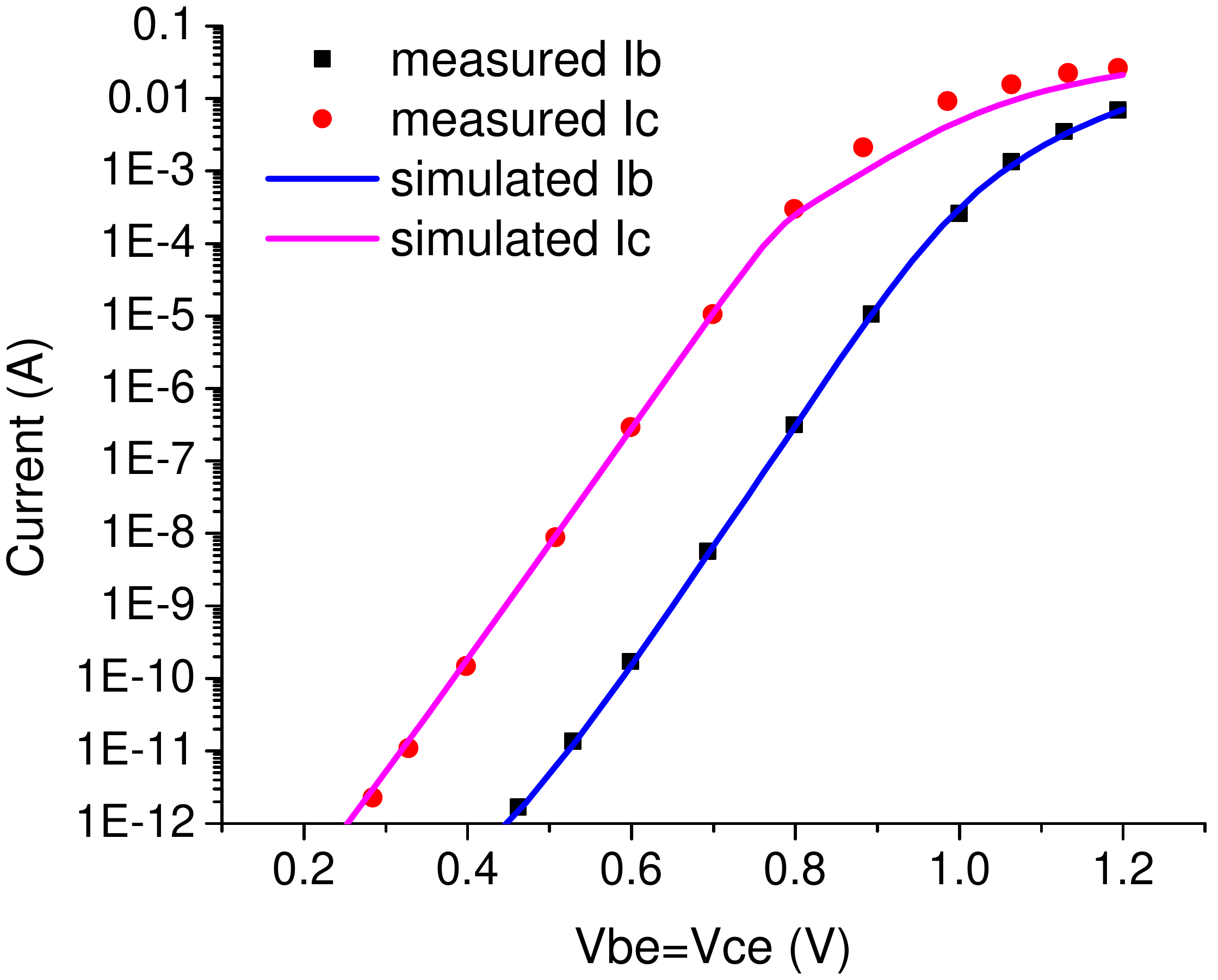}
  \caption{Comparison of the simulated I-V characteristics with the measured data ~\cite{Bock2015}.}
  \label{fig2}
\end{figure}

Fig.~\ref{fig1} shows the structure of the SiGe HBT in Synopsys Sentaurus TCAD. The model is set up to simulate a double-polysilicon self-aligned (DPSA) SiGe HBT with an emitter area of 0.13$\times$2.73\,$\mu$m$^{2}$~\cite{Chevalier2011,Bock2015}. Default material parameters for Si and SiGe in Sentaurus are used for the TCAD model. The model accounts for the physical mechanisms including the hydrodynamic transport, velocity saturation, and generation-recombination~\cite{Sentaurus2017}. For the boundary conditions, the emitter, base, and collector contacts are set as Ohmic contacts. The device parameters in the TCAD model are tuned to fit the measured I-V characteristics~\cite{Bock2015}. Table~\ref{table1} summarizes the adjusted device parameters. Fig.~\ref{fig2} shows good agreement between the simulated results and measured data. 

\begin{table}[t]
\caption{Summary of the device parameters for the SiGe HBT TCAD model.}
\label{table1}       
\begin{center}
\begin{tabular}{|l|c|}
    \hline
    {\bfseries Device parameters} & {\bfseries Value} \\
    \hline
    Emitter width We ($\mu$m) & 0.21 \\
    \hline
    Base width Wb ($\mu$m) & 0.02 \\
    \hline
    Collector width Wc ($\mu$m) & 0.4 \\
    \hline
    Sub-Collector width Wsc ($\mu$m) & 0.2 \\
    \hline
    Emitter length Le ($\mu$m) & 0.13 \\
    \hline
    Base length Lb ($\mu$m) & 0.35 \\
    \hline
    Collector length Lc ($\mu$m) & 0.15 \\
    \hline
    Emitter contact length Lec ($\mu$m) & 0.18 \\
    \hline
    Base contact length Lbc ($\mu$m) & 0.15 \\
    \hline
    Collector contact length Lcc ($\mu$m) & 0.15 \\
    \hline
    SiGe and Base PolySi overlap length Lov ($\mu$m) & 0.05 \\
    \hline 
    Shallow trench isolation (STI) length Lox ($\mu$m) & 0.2 \\
    \hline 
    PolySi doping (cm$^{-3}$) & 1e20 \\
    \hline 
    Base doping (cm$^{-3}$) & 3e18 \\
    \hline 
    Selectively Implanted Collector (SIC) doping (cm$^{-3}$) & 1e16 \\
    \hline 
    Collector doping (cm$^{-3}$) & 1e15 \\
    \hline 
    Sub-collector doping (cm$^{-3}$) & 2e19 \\
    \hline 
\end{tabular}
\end{center}
\end{table}

\begin{figure}[ht]
  \centering
  \includegraphics[width=\columnwidth]{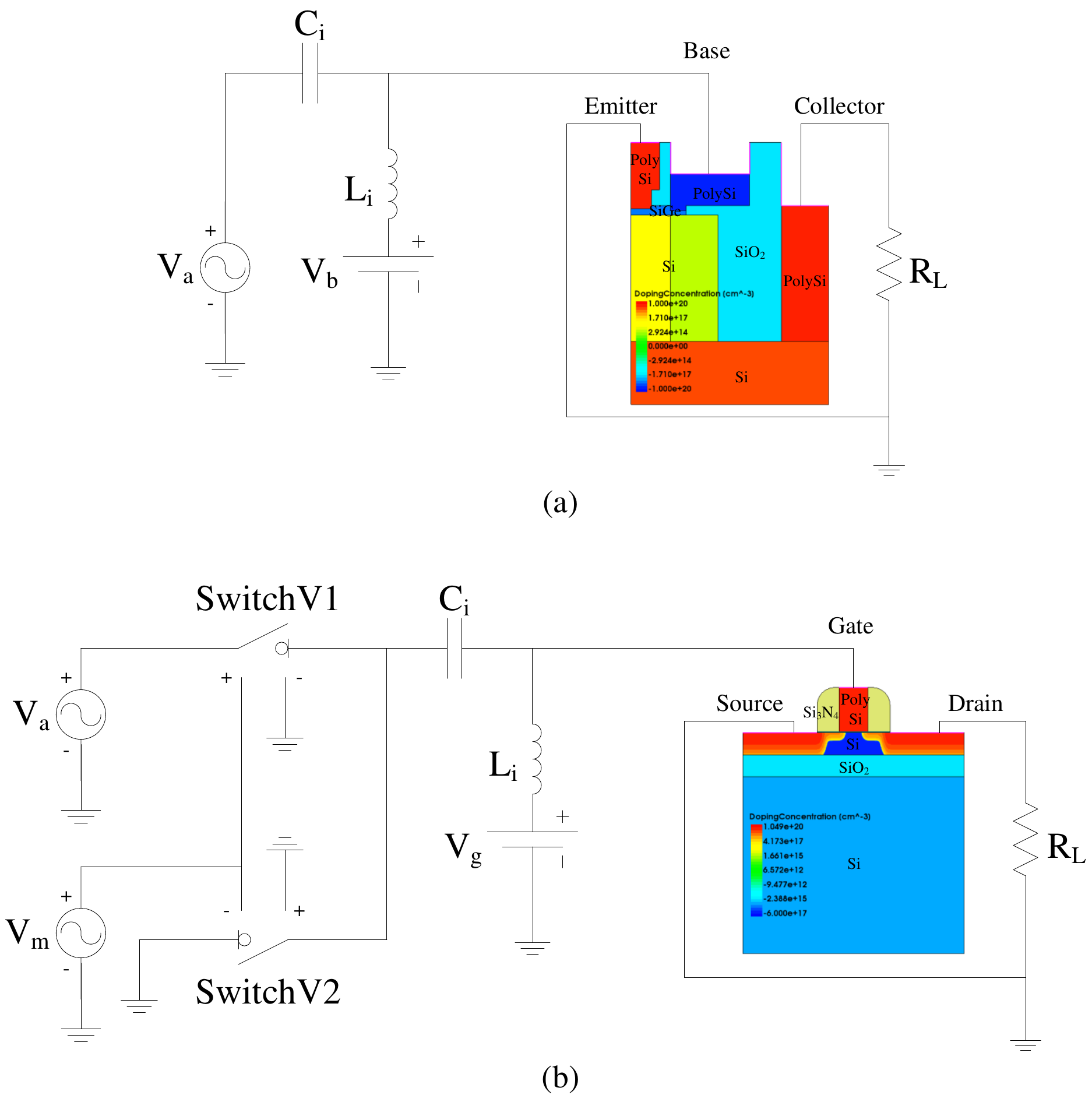}
  \caption{Schematics of mixed mode simulation for THz detection with the TCAD models (a) without modulation and (b) with modulation.}
  \label{fig3}
\end{figure}

\begin{figure}[ht]
  \centering
  \includegraphics[width=0.75\columnwidth]{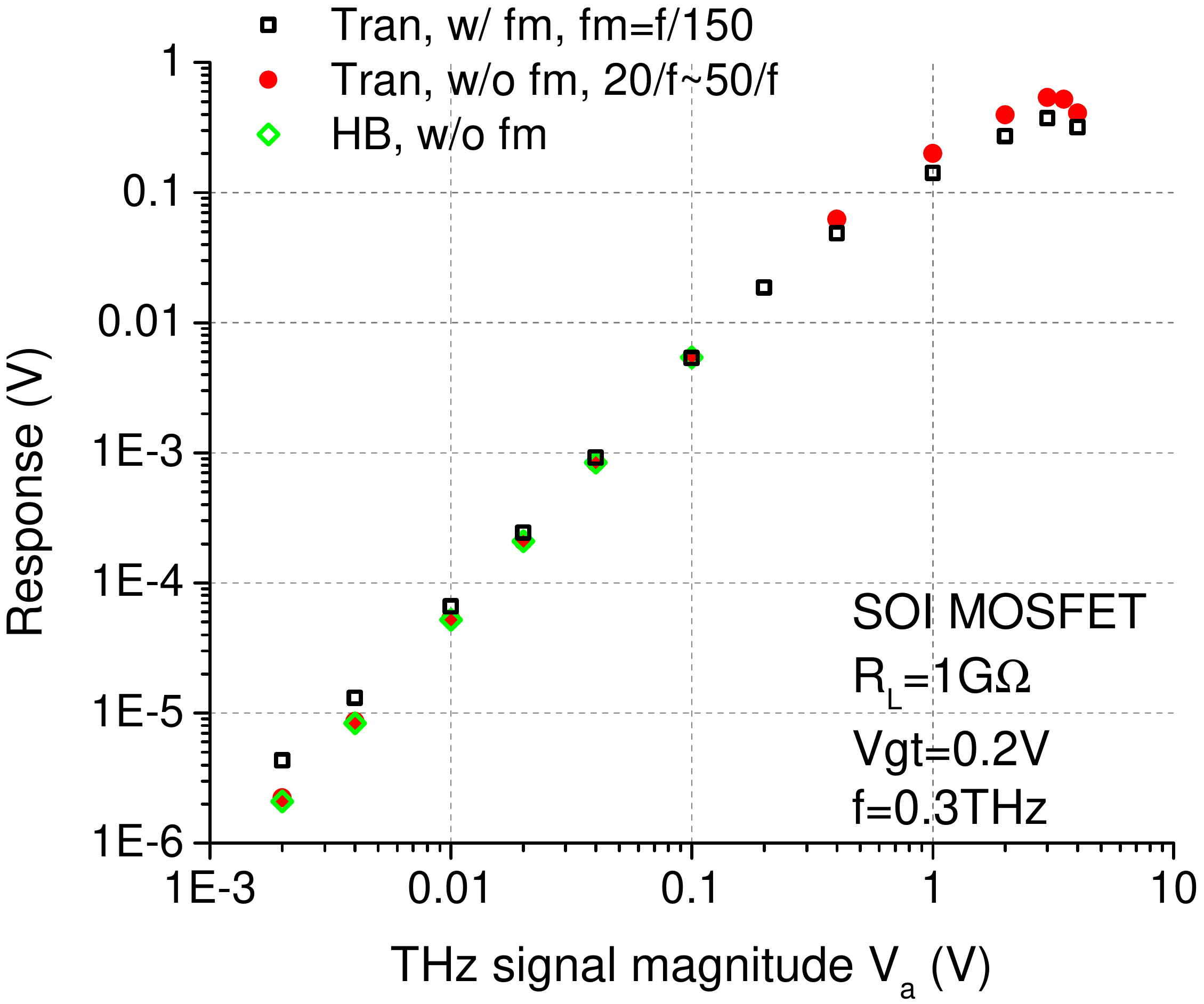}
  \caption{Simulated drain response at 0.3\,THz as a function of the THz signal magnitude for the SOI MOSFET TCAD model with different schemes.}
  \label{fig4}
\end{figure}

\begin{figure}[ht]
  \centering
  \includegraphics[width=0.75\columnwidth]{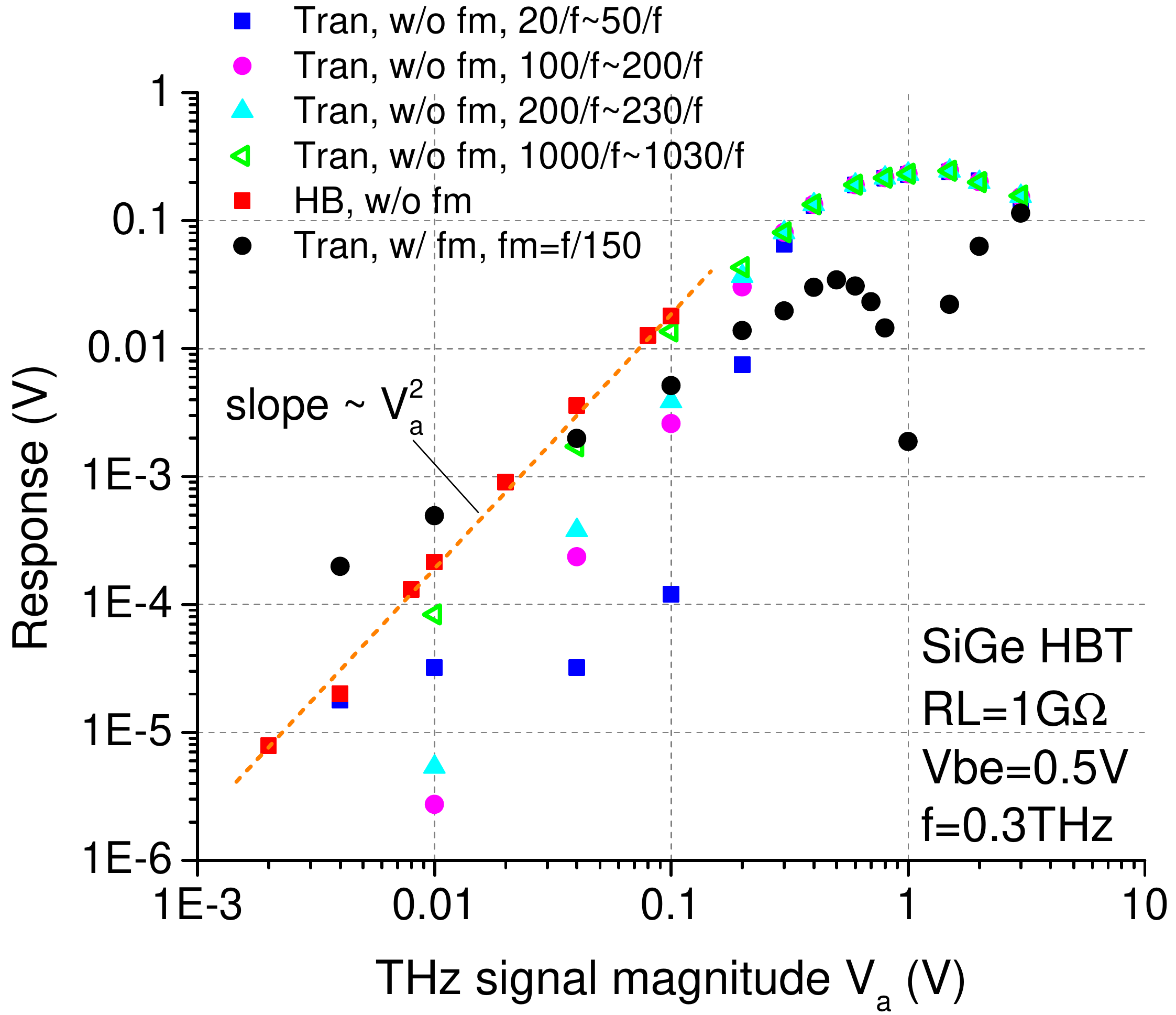}
  \caption{Simulated collector response at 0.3\,THz as a function of the THz signal magnitude for the SiGe HBT TCAD model with different schemes.}
  \label{fig5}
\end{figure}

\section{Response simulation approach}

For the modeling of THz detection, we ran mixed mode simulations in Sentaurus with the TCAD model. We also include the validated TeraFET TCAD models for comparison~\cite{Liu201901,Liu2019}. Fig.~\ref{fig3} shows the two schemes of the mixed mode simulations in Sentaurus. The schematic in Fig.~\ref{fig3} (a) is the typical configuration illustrated for theoretical analysis of the THz response, where the THz signal is not modulated. This configuration could be simulated with either transient or harmonic balance (HB) simulations in Sentaurus. For the transient simulation, the THz voltage response at the collector or the drain could be extracted from the Fourier transform of the collector or drain voltage waveform at the zero frequency or dc component. For the HB simulation, the THz voltage response could be simply extracted from the zero or dc component of the collector or drain voltage when the simulation is completed. The schematic in Fig.~\ref{fig3} (b) is the typical configuration used in practical measurements, where the THz signal $V_{a}$ is modulated by another ac voltage source $V_{m}$ with a modulation frequency $f_{m}$ much lower than the THz signal frequency $f$. This configuration could only be simulated with the time-varying transient simulation in Sentaurus. The THz response could also be extracted from the Fourier transform of the collector or drain voltage waveform, but at the modulation frequency $f_{m}$.

The time-varying schemes with or without modulation have been validated in~\cite{Liu2019}. To evaluate the HB scheme, we use the SOI MOSFET TeraFET TCAD model in~\cite{Liu2019}. Fig.~\ref{fig4} shows the comparison of the simulated drain response as a function of the THz signal magnitude for the SOI MOSFET model. The response of the transient simulation without modulation is extracted from the waveform with the time range between $20/f$ and $50/f$. The transient simulation results with and without modulation are in good agreement and the results with modulation deviate from the quadratic response (proportional to $V_a^2$) due to the non-ideal voltage-controlled switches in Fig.~\ref{fig3} (b). This is consistent with the results of the AlGaAs/InGaAs HFET in~\cite{Liu2019}. Also, the HB simulation results agree with the transient simulation results without modulation, which validates the HB approach for THz response simulation in Sentaurus. At high THz signal magnitude, the HB simulation suffers from convergence problems and transient simulation could be used.

The different schemes are then applied to the SiGe HBT TCAD simulations. Fig.~\ref{fig5} shows the dependence of the simulated response on the THz signal magnitude for the SiGe HBT model. For the transient simulation scheme without modulation, different time ranges for the collector voltage waveform are used for the extraction of the THz response. The comparison shows that the time-varying simulation with modulation may not be proper for the SiGe HBT response simulation, since the HBT may not have large enough input impedance to minimize the influence of the non-ideal voltage-controlled switches. Also, the time range of the collector voltage waveform for the THz response extraction from the Fourier transform could have a significant effect on the results at the low or medium $V_a$ ranges. It could be seen that the longer time the transient simulation takes, the more accurate results are obtained since they are getting closer to the HB simulation results which could be considered as the correct results. At high intensities ($V_a>$ 0.2\,V), however, the results are the same for different time ranges used for the response extraction. This suggests an efficient simulation approach for the SiGe HBT response, which is to use HB simulations in the low and intermediate $V_a$ ranges and transient simulation without modulation in the high $V_a$ range. The results also show the quadratic response at the low and medium intensity range and the response saturation at the high intensity range for the SiGe HBT, which is consistent with the $V_a$ dependence of the response for the TeraFET detectors~\cite{Liu2019}.

\begin{figure}[t]
  \centering
  \includegraphics[width=0.75\columnwidth]{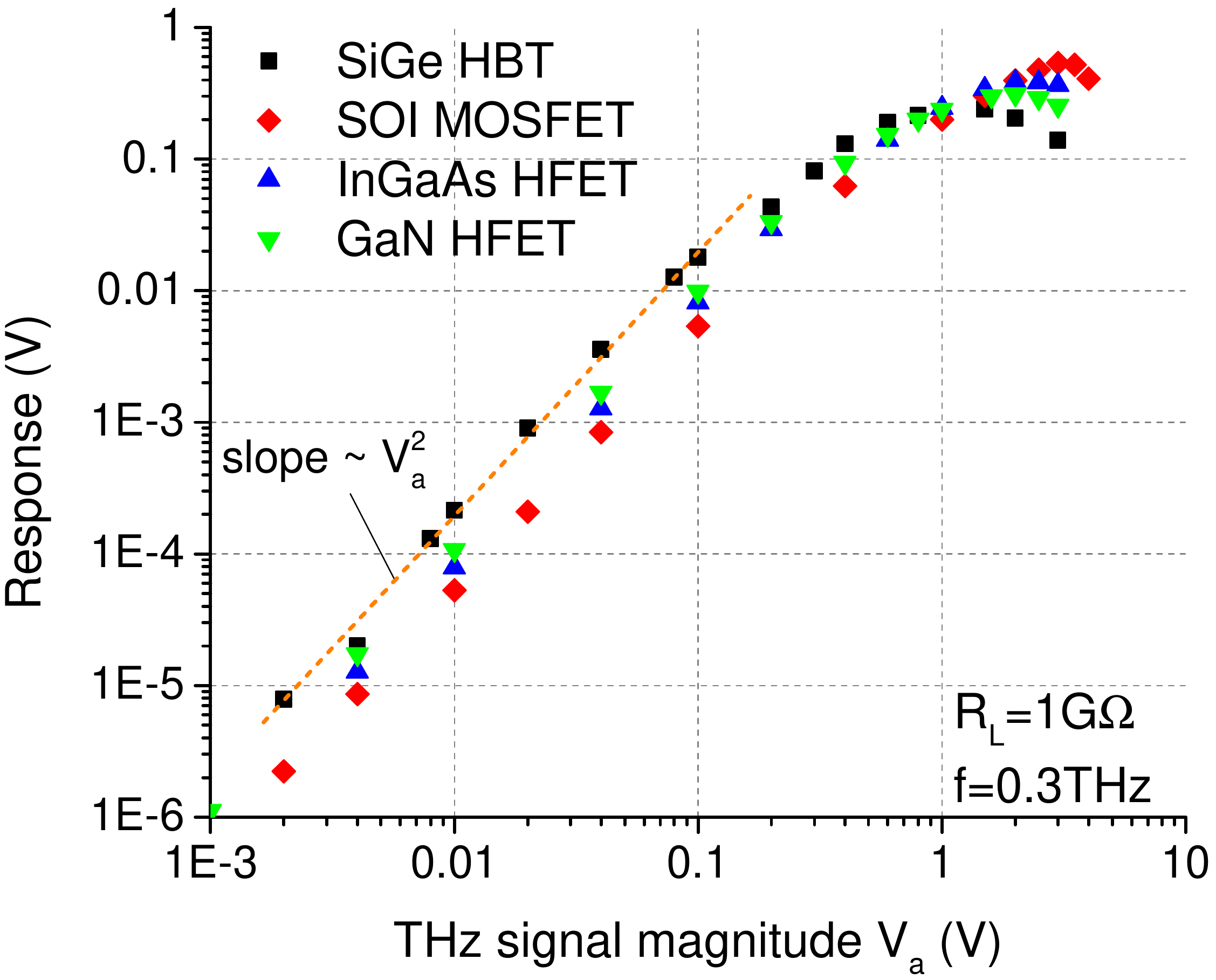}
  \caption{Comparison of simulated collector or drain response at 0.3\,THz as a function of the THz signal magnitude for the SiGe HBT and TeraFET TCAD models.}
  \label{fig6}
\end{figure}

\section{Comparison with THz TeraFET detectors}

Using the proposed scheme, we obtain the dynamic range of the THz response for the SiGe HBT model and compare the results with the TeraFET TCAD models in~\cite{Liu2019}. Fig.~\ref{fig6} shows the comparison of the dynamic ranges. The results are obtained using the transient simulation without modulation, except for the SiGe HBT model in the 1\,mV and 0.1\,V range of the THz signal magnitude; where the response is obtained from the HB simulation. It could be seen that these devices have comparable dynamic ranges, and SiGe HBTs could work as efficient THz detectors as the TeraFETs implemented in different material systems.

\begin{figure}[t]
  \centering
  \includegraphics[width=\columnwidth]{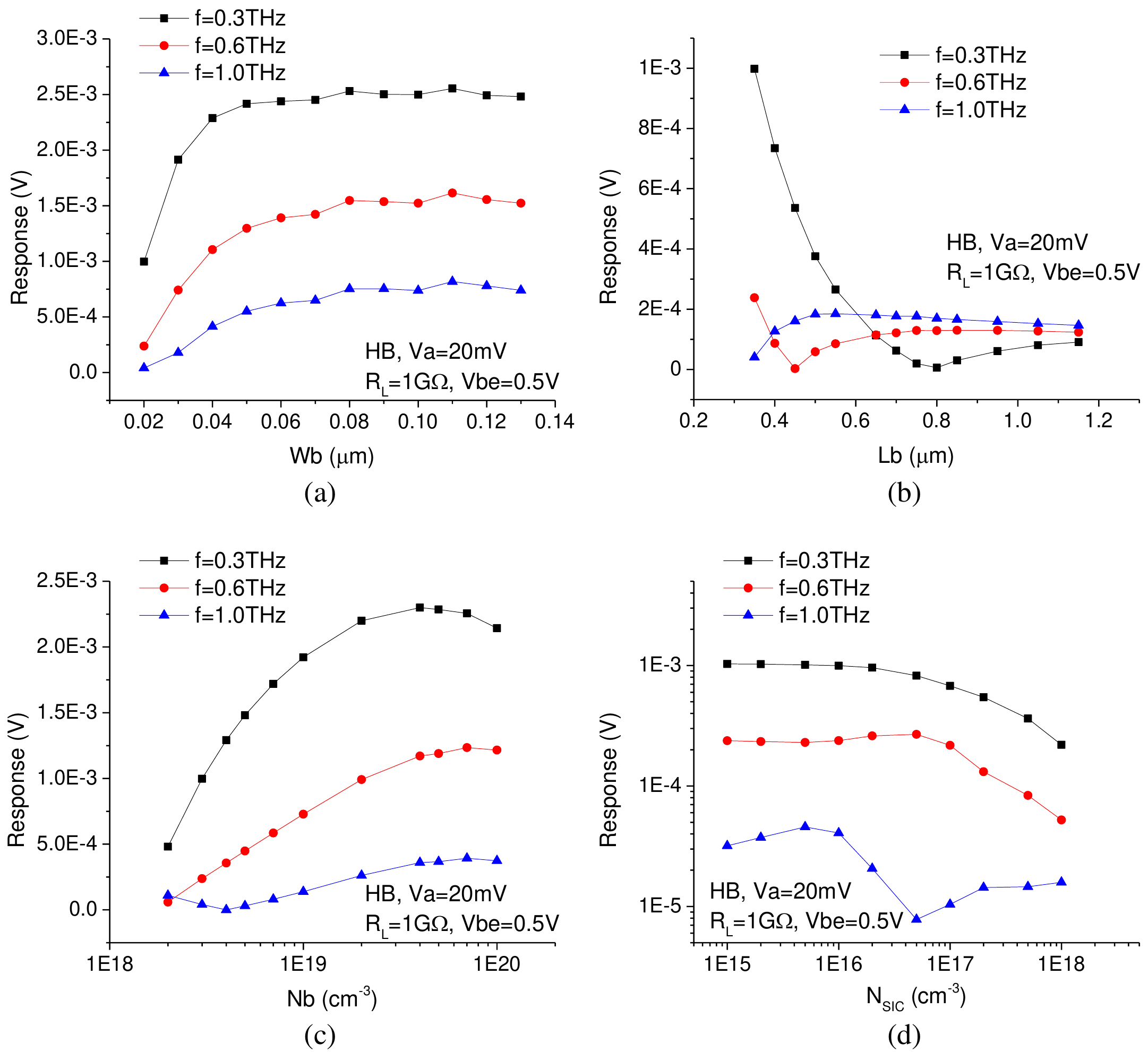}
  \caption{Simulated collector response to the variations of the device design parameters for the SiGe HBT TCAD model at different frequencies.}
  \label{fig7}
\end{figure}

\section{Response to design parameter variation}

The mixed mode HB simulation could be a very useful tool for the physical design of the THz detector, since it runs much faster than the time-varying transient simulation. By running the HB simulation at the intermediate Va range, we explore the SiGe HBT response to the variations of the device design parameters at different frequencies. Fig.~\ref{fig7} shows the simulated results for various base parameter values including the width, length, and doping, as well as the collector doping values. Therefore, high detector response could be obtained by optimizing the device parameters, such as adopting medium base width, small base length, high base doping and low SIC doping. These results also suggest an effective approach of using THz scanning to test Si BiCMOS VLSI, since the THz response could be different between the working and defective devices or the authentic and counterfeit chips.

\section{Conclusion}

The TCAD model for the SiGe HBT detector is in good agreement with the measured current-voltage characteristics and shows the quadratic response at the low and medium intensity ranges and the response saturation at the high intensity range. The dynamic range is comparable with the dynamic range of the various TeraFET detectors. The mixed mode harmonic balance simulation for the SiGe HBT response to the variations of the device parameters provides an efficient approach for the detector design and optimization. The simulation results could be applied in the non-destructive testing of Si BiCMOS VLSI with THz radiation.


\begin{thebibliography}{99}

\bibitem{Akkas201901} M. A. Akka{\c{s}}, ``Terahertz wireless data communication,'' \emph{Wireless Netw.}, \textbf{25}, January 2019, pp. 145--155.

\bibitem{Klatt200912} G. Klatt \textit{et al.}, ``Rapid-scanning terahertz precision spectrometer with more than 6 THz spectral coverage,'' \emph{Opt. Express}, \textbf{17}, 25, December 2009, pp. 22847--22854.

\bibitem{Federici200506} J. F. Federici \textit{et al.}, ``THz imaging and sensing for security applications-—explosives, weapons and drugs,'' \emph{Semicond. Sci. Technol.}, \textbf{20}, 7, June 2005, pp. S266--S280.

\bibitem{Cheon201611} H. Cheon \textit{et al.}, ``Terahertz molecular resonance of cancer DNA,'' \emph{Sci. Rep.}, \textbf{6}, November 2016, Art. No. 37103.

\bibitem{Xie201503} L. Xie \textit{et al.}, ``Extraordinary sensitivity enhancement by metasurfaces in terahertz detection of antibiotics,'' \emph{Sci. Rep.}, \textbf{5}, March 2015, Art. No. 8671.

\bibitem{Ahi201803} K. Ahi, S. Shahbazmohamadi, and N. Asadizanjani, ``Quality control and authentication of packaged integrated circuits using enhanced-spatial-resolution terahertz time-domain spectroscopy and imaging,'' \emph{Opt. Lasers Eng.}, \textbf{104}, May 2018, pp. 274--284.

\bibitem{Dyakonov199605} M. I. Dyakonov and M. S. Shur, ``Detection, mixing, and frequency multiplication of terahertz radiation by two-dimensional electronic fluid,'' \emph{IEEE Trans. Electron Devices}, \textbf{43}, 3, March 1996, pp. 380--387. 

\bibitem{Dyakonov199610} M. I. Dyakonov and M. S. Shur, ``Plasma wave electronics: novel terahertz devices using two-dimensional electron fluid,'' \emph{IEEE Trans. Electron Devices}, \textbf{43}, 10, October 1996, pp. 1640--1645.

\bibitem{Popov201104} V. Popov \textit{et al.}, ``High-responsivity terahertz detection by on-chip InGaAs/GaAs field-effect-transistor array,'' \emph{Appl. Phys. Lett.}, \textbf{98}, 15, April 2011.

\bibitem{Gavrilenko200702} V. Gavrilenko \textit{et al.}, ``Electron transport and detection of terahertz radiation in a GaN/AlGaN submicrometer field-effect transistor,'' \emph{Semicond.}, \textbf{41}, 2, February 2007, pp. 232--234.

\bibitem{Knap200407} W. Knap \textit{et al.}, ``Plasma wave detection of sub-terahertz and terahertz radiation by silicon field-effect transistors,'' \emph{Appl. Phys. Lett.}, \textbf{85}, 4, July 2004, pp. 675--677.

\bibitem{Coquillat201612} D. Coquillat \textit{et al.}, ``High-speed room temperature terahertz detectors based on InP double heterojunction bipolar transistors,'' \emph{Int. J. High Speed Electron. Syst.}, \textbf{25}, 03n04, December 2016, Art. No. 1640011.

\bibitem{Dyakonova2017} N. Dyakonova \textit{et al.}, ``Detection of high intensity THz radiation by InP double heterojunction bipolar transistors,'' in \emph{2017 42nd Int. Conf. Infrared, Millimeter, Terahertz Waves}, 2017, pp. 1--2.

\bibitem{Mai201706} A. Mai \textit{et al.}, ``High-Speed SiGe BiCMOS Technologies and Circuits,'' \emph{Int. J. High Speed Electron. Syst.}, \textbf{26}, 01n02, June 2017, Art. No. 1740002.

\bibitem{Schroter201601} M. Schr{\"o}ter \textit{et al.}, ``SiGe HBT technology: Future trends and TCAD-based roadmap,'' \emph{Proceedings of the IEEE}, \textbf{105}, 6, January 2016, pp. 1068--1086.

\bibitem{Hadi2012} R. Al Hadi, J. Grzyb, B. Heinemann, and U. Pfeiffer, ``Terahertz detector arrays in a high-performance SiGe HBT technology,'' in \emph{IEEE Bipolar/BiCMOS Circuits and Technology Meeting}, 2012, pp. 1--4.

\bibitem{Ghodsi201906} H. Ghodsi and H. Kaatuzian, ``Analysis and design of a SiGe-HBT based terahertz detector for imaging arrays applications,'' \emph{Microelectronics J.}, \textbf{90}, June 2019, pp. 156--162.

\bibitem{Chevalier2011} P. Chevalier \textit{et al.}, ``Towards THz SiGe HBTs,'' in \emph{IEEE Bipolar/BiCMOS Circuits and Technology Meeting}, 2011, pp. 57--65.

\bibitem{Bock2015} J. B{\"o}ck \textit{et al.}, ``SiGe HBT and BiCMOS process integration optimization within the DOTSEVEN project,'' in \emph{IEEE Bipolar/BiCMOS Circuits and Technology Meeting}, 2015, pp. 121--124.

\bibitem{Sentaurus2017} Sentaurus Device User Guide (Version N-2017.09), Synopsys Inc., Mountain View, CA, USA, 2017.

\bibitem{Liu201901} X. Liu and M. Shur, ``An efficient TCAD model for TeraFET detectors,'' in \emph{IEEE Radio and Wireless Symp.}, 2019, pp. 1--4.

\bibitem{Liu2019} X. Liu and M. S. Shur, ``TCAD model for TeraFET detectors operating in a large dynamic range,'' \emph{IEEE Trans. Terahertz Sci. Technol.}, to be published.

\end{thebibliography}
\end{document}